\documentclass[aps,prl,amssymb,twocolumn,superscriptaddress,10pt,tightenlines]{revtex4-1}

\usepackage{graphicx}
\usepackage{dcolumn}
\usepackage{bm}
\usepackage{color}
\usepackage{subfigure}

\begin{document}

\newcommand{\beq}{\begin{equation}}
\newcommand{\eeq}{\end{equation}}
\newcommand{\barr}{\begin{eqnarray}}
\newcommand{\earr}{\end{eqnarray}}

\def\bra#1{\langle{#1}|}
\def\ket#1{|{#1}\rangle}
\def\sinc{\mathop{\text{sinc}}\nolimits}
\def\cV{\mathcal{V}}
\def\cH{\mathcal{H}}
\def\cT{\mathcal{T}}
\renewcommand{\Re}{\mathop{\text{Re}}\nolimits}
\newcommand{\tr}{\mathop{\text{Tr}}\nolimits}

\definecolor{dgreen}{rgb}{0,0.5,0}
\newcommand{\green}{\color{dgreen}}
\newcommand{\BLUE}[1]{\text{\color{blue} #1}}
\newcommand{\GREEN}[1]{\textbf{\color{green}#1}}
\newcommand{\REV}[1]{\textbf{\color{red}[[#1]]}}
\newcommand{\KY}[1]{\textbf{\color{dgreen}[[#1]]}}
\newcommand{\rev}[1]{{\color{red}[[#1]]}}

\def\HN#1{{\color{magenta}#1}}
\def\DEL#1{{\color{yellow}#1}}

\title{Correlation plenoptic imaging}

\author{Milena D'Angelo}\email{milena.dangelo@uniba.it}
\affiliation{Dipartimento Interateneo di Fisica, Universit\`a degli studi di Bari, I-70126 Bari, Italy} 
\affiliation{INFN, Sezione di Bari, I-70126 Bari, Italy}

\author{Francesco V. Pepe}\email{francesco.pepe@ba.infn.it}
\affiliation{Museo Storico della Fisica e Centro Studi e Ricerche ``Enrico Fermi'', I-00184 Roma, Italy} 
\affiliation{INFN, Sezione di Bari, I-70126 Bari, Italy}

\author{Augusto Garuccio}
\affiliation{Dipartimento Interateneo di Fisica, Universit\`a degli studi di Bari, I-70126 Bari, Italy} 
\affiliation{INFN, Sezione di Bari, I-70126 Bari, Italy}

\author{Giuliano Scarcelli}
\affiliation{Fischell Department of Bioengineering, University of Maryland, College Park MD 20742 USA}

\begin{abstract}
\noindent Plenoptic imaging is a promising optical modality that simultaneously captures the  location and the propagation direction of light in order to enable three-dimensional imaging in a  single shot. However, in standard plenoptic imaging systems, the maximum spatial and angular resolutions are fundamentally linked; thereby, the maximum achievable depth of field is inversely proportional to the spatial resolution. We propose to take advantage of the second-order correlation properties of light to overcome this fundamental limitation. In this paper, we demonstrate that the correlation in both momentum and position of chaotic light leads to the enhanced refocusing power of correlation plenoptic imaging with respect to standard plenoptic imaging.
\end{abstract}

\maketitle

\begin{figure}
\centering
{\includegraphics[width=0.48\textwidth]{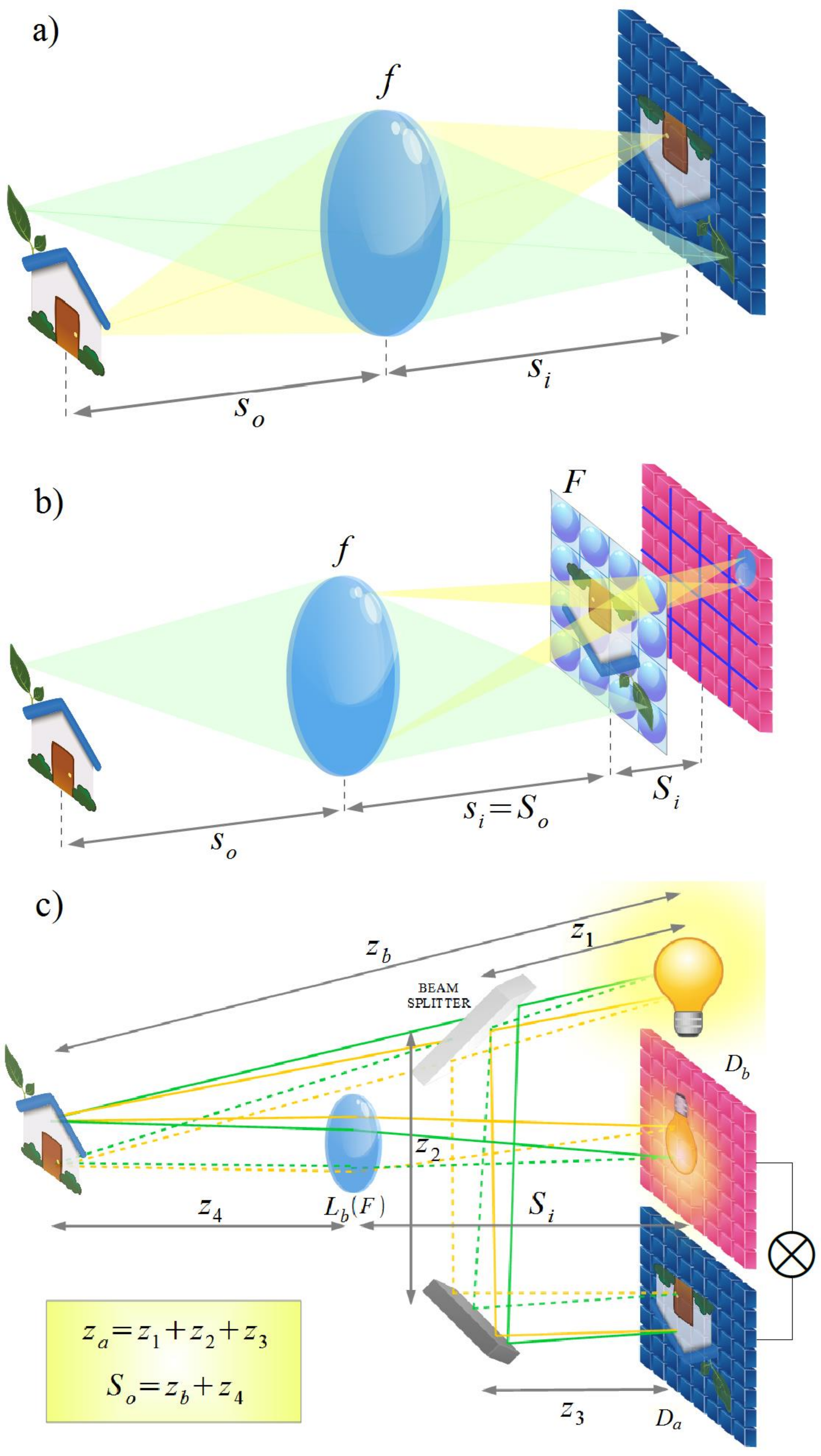}}
\caption{a) Standard imaging: The lens focuses the object on the sensor. 
b) Plenoptic imaging: The lens focuses the object on a lenslet array; each microlens focuses the main lens on a \textit{macropixel} (blue lines) of the sensor to provide directional information. 
c) Correlation plenoptic imaging: By means of correlation measurement, the chaotic light source acts as a focusing element (i.e., as an imaging lens) and enables imaging the object on the (blue) sensor $\mathrm{D}_a$. The light source is imaged by the lens $L_b$ on the (red) sensor $\mathrm{D}_b$ to provide directional information. Note: Distances are related by the thin-lens equations: $1/s_0+1/s_i=1/f$, $1/S_0+1/S_i=1/F$. }\label{fig:schemes}
\end{figure}

\noindent Plenoptic imaging (PI) is a technique aimed at capturing information on the three-dimensional lightfield of a given scene in a single shot \cite{adelson}. Its key principle is to record, in the image plane, both the location and the propagation direction of the incoming light. The recorded propagation direction is exploited, in post-processing, to computationally retrace the geometrical light path, thus enabling the refocusing of different planes within the scene and the extension of the depth of field of the acquired image. As shown in Fig.\ref{fig:schemes}b, PI resembles standard imaging (Fig.\ref{fig:schemes}a), however, a microlens array is inserted in the native image plane and the sensor array is moved behind the microlenses. On the one hand, the microlenses act as imaging pixels to gain the spatial information of the scene; on the other hand, each microlens reproduces on the sensor array an image of the camera lens, thus providing the angular information associated with each imaging pixel \cite{ng}. As a result, a trade-off between spatial and angular resolution is built in the plenoptic imaging process.

Plenoptic imaging is currently used in digital cameras enhanced by refocusing capabilities \cite{website}; in fact, in photography, PI highly simplifies both auto-focus and low-light shooting \cite{ng}. A plethora of innovative applications in 3-D imaging and sensing \cite{3dimaging,waller_turb}, stereoscopy \cite{adelson,muenzel,levoy} and microscopy \cite{microscopy1,microscopy2,microscopy3} are also being developed. In particular, high-speed large-scale 3D funcional imaging of neuronal activity has been demonstrated \cite{microscopy4}. However, the potentials of PI are strongly limited by the inherent inverse proportionality between image resolution and maximum achievable depth of field. Attempts to decouple resolution and depth of field based on signal processing and deconvolution have been proposed in literature \cite{microscopy2,microscopy4,waller,spatioangular,imageformation,superres}. 

Our idea is to exploit the second-order spatio-temporal correlation properties of light to overcome this fundamental limitation. Using two correlated beams, from either a chaotic or an entangled photon source, we can perform imaging in one arm \cite{pittman,gatti2,laserphys,valencia,scarcelliPRL,ferri}, and simultaneously obtain the angular information in the other arm. In fact, the position and momentum correlations at the core of our proposal have been demonstrated more than ten years ago by performing separate imaging and diffraction experiments, respectively \cite{laserphys,ferri}. Here we devise a physical context where such correlations can be measured and exploited simultaneously to enhance the performances of a practically-useful imaging technique, namely, to improve the depth of field of plenoptic imaging. In this paper, we develop a comprehensive theory of the proposed techinque, named correlation plenoptic imaging (CPI), in the case of chaotic light. In particular, we show that the second order correlation function possesses plenoptic imaging properties (i.e., it encodes both spatial and angular information), and is thus characterized by a key re-focusing capability. From a practical standpoint, our protocol can dramatically enhance the potentials of PI, the simplest method of 3D imaging with the present technological means \cite{microscopy2,microscopy4}. From a fundamental standpoint, the plenoptic application is the first situation where the counterintuitive properties of correlated systems are effectively used to beat intrinsic limits of standard imaging systems. Moreover, the interest in position-momentum correlations goes well beyond imaging and optics, due to their relation with quantum tomography \cite{mancini,lvovsky}.

The working principle of CPI is introduced in Fig. \ref{fig:schemes}c. In PI (Fig.~\ref{fig:schemes}b) the sensor is divided into {\it macropixels} of width $\delta_x$, defining the image resolution; a macropixel is made of $N_u^\mathrm{(p)}$ micropixels per side, of width $\delta<\delta_x$, fixing the directional resolution \cite{adelson,ng}. An array of microlenses of diameter $\delta_x$ and focal length $F$ is inserted in front of the sensor for reproducing, within each macropixel, the image of the main camera lens.
Hence, each micropixel collects light from a sector of the main lens and encodes information on the direction of light that impinges on the specific macropixel, corresponding to a specific point of the acquired image. For a sensor of width $W$, this configuration yields the following relationship between the number of pixels per side devoted to the spatial ($N_x^\mathrm{(p)}=W/\delta_x$) and to the directional ($N_u^\mathrm{(p)}=\delta_x/\delta$) detection of the lightfield:
\begin{equation}\label{Np}
N_x^{\mathrm{(p)}} N_u^{\mathrm{(p)}} = \frac{W}{\delta} \equiv
N_{\mathrm{tot}},
\end{equation}
where $N_{\mathrm{tot}}$ is the total number of pixels per side on the sensor. Simple geometrical considerations indicate that the maximum achievable depth of field is determined by the directional resolution $N_u^\mathrm{(p)}$; hence, based on Eq.~(\ref{Np}), the depth of field can be increased only at the expenses of the image resolution $N_x^\mathrm{(p)}$. In addition, for the direction measurement to be physically meaningful, the image resolution must be well above the diffraction limit, namely, the image pixel size $\delta_x$ must be bigger than the minimum image resolution allowed by diffraction.

\noindent Now, in CPI (Fig.~\ref{fig:schemes}c) light is split in two arms, and two sensors are employed: $\mathrm{D}_a$ to image the desired scene, and $\mathrm{D}_b$ to perform the direction measurement. In fact, when using the correlation properties of chaotic light, the light source plays the role of the focusing element, and can replace the main lens of Figs.~\ref{fig:schemes}a-\ref{fig:schemes}b \cite{valencia,scarcelliPRL}. Direction measurement is thus performed by imaging the light source on $\mathrm{D}_b$ through the lens $L_b$, which collects light reflected off the object. Notice that lens $L_b$ replaces the whole lenslet array of Fig.~\ref{fig:schemes}b. 
As we shall prove, when measuring correlation between $\mathrm{D}_a$ and $\mathrm{D}_b$, the image of the desired scene is reproduced on $\mathrm{D}_a$. The image is focused provided the distance $z_a$ between the source and the sensor $\mathrm{D}_a$ is equal to the distance $z_b$ between the source and the object \cite{scarcelliPRL,note_La}.  
Since the two sensors can have the same pixel size $\delta$, and their widths $W_i^\mathrm{(cp)}$ (with $i=x,u$) are such that $W_x^\mathrm{(cp)}+W_u^\mathrm{(cp)}=W$, the number of pixels per side dedicated to the spatial and the directional measurement ($N_i^\mathrm{(cp)}=W_i^\mathrm{(cp)}/\delta$, with $i=x,u$) is constrained by the relation:
\begin{equation}\label{Nc}
N_x^{\mathrm{(cp)}} + N_u^{\mathrm{(cp)}} =N_{\mathrm{tot}}.
\end{equation}
The striking improvement of CPI with respect to PI increases with increasing $N_{\mathrm{tot}}$. Realistic chaotic sources for the realization of the proposed scheme include the pseudo-thermal sources typically employed in ghost imaging \cite{gatti2,valencia,scarcelliPRL,ferri,torino}, as well as LEDs \cite{note_time}.

\textit{Correlation imaging and refocusing.---}
The core of CPI is the second-order spatio-temporal correlation measurement, as described by Glauber correlation function \cite{scully}
\begin{eqnarray}\label{glauber}
G^{(2)}(\bm{\rho}_a,\bm{\rho}_b;t_a,t_b) & = & \left\langle E^{(-)}_a
(\bm{\rho}_a,t_a) E^{(-)}_b (\bm{\rho}_b,t_b) \right. \nonumber
\\ & & \left. E^{(+)}_b (\bm{\rho}_b,t_b) E^{(+)}_a
(\bm{\rho}_a,t_a) \right\rangle,
\end{eqnarray}
where $\bm{\rho}_{i}$ indicates the planar position on the sensor $\mathrm{D}_{i}$ (with $i=a,b$), $t_{i}$ is the corresponding detection time, and $E_{i}^{(\pm)}$ are the positive- and negative-frequency components of the electric field operators at each detector, for which a scalar approximation is assumed. The expectation value in Eq.(\ref{glauber}) is evaluated by considering the source statistics. Let us consider a quasi-monochromatic chaotic light source characterized by a coherence time larger than $\tau=t_a-t_b$, in such a way that the temporal part of the correlation function can be neglected, and the spatial part reduces to \cite{laserphys}: 
\begin{equation}\label{G2}
G^{(2)}(\bm{\rho}_a,\bm{\rho}_b) = I_{a} (\bm{\rho}_a) I_{b} (\bm{\rho}_b) + \Gamma (\bm{\rho}_a,\bm{\rho}_b),
\end{equation}
where the first term is the mere product of the intensities $I_{i}=G_{ii}^{(1)}$ at the two detectors ($i=a,b$), and the second term is the non trivial part of the correlation $\Gamma=|G_{ab}^{(1)}|^2$. Here, $G_{jk}^{(1)} (\bm{\rho}_j,\bm{\rho}_k) = C \! \int
d^2\bm{q}\, g_j^*(\bm{\rho}_j,\bm{q})
g_k(\bm{\rho}_k,\bm{q})$, with $C$ a constant and $g_{i}(\bm{\rho}_{i},\bm{q})$ the Green's function propagating the electric field mode with transverse momentum $\bm{q}$ from the source to the detector $\mathrm{D}_{i}$ \cite{goodman}. By propagating the field in the setup of Fig.\ref{fig:schemes}, while assuming for simplicity that lens $L_b$ is diffraction-limited, the second term of Eq.(\ref{G2}) yields the desired correlation plenoptic image of the object:
\begin{eqnarray}\label{G12}
& & \Gamma_{z_a,z_b} (\bm{\rho}_a,\bm{\rho}_b) = C' \left| \int d^2 \bm{\rho}_o
A(\bm{\rho}_0) e^{-\frac{i\omega}{c z_b} \bm{\rho}_o \cdot
\frac{\bm{\rho}_b}{M} } \right. \nonumber \\ & & \left. \times \int d^2 \bm{\rho}_s F(\bm{\rho}_s)
G(|\bm{\rho}_s|)_{\left[\frac{\omega}{c}\!\left(\frac{1}{z_b}-
\frac{1}{z_a} \right)\!\right]} e^{-\frac{i\omega}{c z_a} \!\left(\frac{z_a}{z_b}
\bm{\rho}_o - \bm{\rho}_a \right)\! \cdot
\bm{\rho}_s} \right|^2, \nonumber \\
\end{eqnarray}
parametrized by the distances of the sensor $\mathrm{D}_a$ ($z_a$) and the object ($z_b$). In Eq.~(\ref{G12}), $C'$ is a constant, $\bm{\rho}_o$ and $\bm{\rho}_s$ are, respectively, the transverse coordinates on the object and the source plane, $A(\bm{\rho}_o)$ is the aperture function describing the object, $F(\bm{\rho}_s)$ is the intensity profile of the source, $G(x)_{[y]} = e^{i y x^2/2}$, and $M=S_i/S_o$ is the magnification of the image of the source on $\mathrm{D}_{b}$. 
Based on the result of Eq.(\ref{G12}), when the sensor $\mathrm{D}_a$ is placed at $z_a=z_b$, for any pixel of the sensor $\mathrm{D}_b$, we obtain
\begin{equation}\label{coherent}
\Gamma_{z_b,z_b} (\bm{\rho}_a,\bm{\rho}_b) \! \propto \! \left| \int\! d^2\! \bm{\rho}_o
A(\bm{\rho}_0) e^{-\frac{i\omega \bm{\rho}_o \cdot \bm{\rho}_b}{c z_b M} }  \tilde{F}\!\left( \frac{\omega(\bm{\rho}_o\!-\!\bm{\rho}_a)}{c z_b} \!\right) \right|^2\!.
\end{equation}
This result indicates that a {\it coherent} image of the object $A(\bm{\rho}_0)$ is retrieved on $\mathrm{D}_a$ when measuring correlation with any pixel of $\mathrm{D}_b$; its point-spread function is given by the Fourier transform of the source intensity profile $\tilde{F}(\bm{\kappa}) = \int d^2 \bm{\rho}_s F(\bm{\rho}_s) e^{-i \bm{\kappa} \cdot \bm{\rho}_s}$. By keeping $z_a=z_b$ and integrating the result of Eq.~(\ref{coherent}) over the whole detector array $\mathrm{D}_b$, one gets the {\it incoherent} image of the object, namely
\begin{eqnarray}\label{incoherent}
& & \Sigma_{z_b}(\bm{\rho}_a) := \int d^2 \bm{\rho}_b
\Gamma_{z_b,z_b}(\bm{\rho}_a,\bm{\rho}_b) \nonumber \\
& & \propto \int d^2 \bm{\rho}_o |A(\bm{\rho}_o)|^2
\!\left| \tilde{F}\!\left( \frac{\omega}{c
z_b}(\bm{\rho}_o-\bm{\rho}_a) \right)\!\right|^2.
\end{eqnarray}
This is the well known ghost image produced by chaotic light sources \cite{valencia,scarcelliPRL}. 
In addition, based on Eq.~(\ref{G12}), the correlation measurement between $\mathrm{D}_b$ and any pixel of $\mathrm{D}_a$ reproduces the image of the source; its point-spread function is given by the Fourier transform of the object aperture function $\tilde{A}(\bm{q}) = \int d^2 \bm{\rho}_o A(\bm{\rho}_o) e^{-i \bm{q}\cdot\bm{\rho}_0}$, which is
\begin{equation}\label{point}
\Gamma_{z_b,z_b}(\bm{\rho}_a,\bm{\rho}_b)\Bigr|_{\text{point source}} \propto
\!\left| \tilde{A} \!\left( \frac{\omega}{c z_b} \left(
\bar{\bm{\rho}}_s + \frac{\bm{\rho}_b}{M} \right) \right)\!
\right|^2
\end{equation}
for a point source placed in $\bar{\bm{\rho}}_s$. Hence, the one-to-one correspondence between points of the source ($\bm{\rho}_s$) and points of the sensor $\mathrm{D}_b$ ($\bm{\rho}_b=-M \bm{\rho}_s$) can only be hindered by diffraction at the object. 

In summary, due to the peculiar position and momentum correlation of chaotic sources, the second order correlation function of Eq.~(\ref{G12}) possesses plenoptic properties, namely, it enables the simultaneous measurement of both spatial and angular information. This intriguing result indicates that plenoptic imaging may represent a natural playground for the position and momentum correlations of chaotic sources to find a realistic practical application. In this perspective, it is worth emphasizing that, based on the result of Eq.(\ref{point}), our scheme may perform plenoptic imaging only when working in the geometrical optics limit. This limit is recovered in both arms of the CPI system when $\lambda z_a /(d D_s) \ll 1 $ with $d$ the smallest detail of the object and $D_s$ the width of the source. It is also worth noticing that the diffraction limits on the image and the angular resolution:
\begin{equation}\label{deltarhoa}
\Delta\rho_a^{\lim} \sim \frac{\lambda z_a}{D_s}, \qquad \frac{\Delta\rho_b^{\lim}}{M} \sim \frac{\lambda z_b}{d},
\end{equation}
are defined, respectively, by the characteristic size of the source and the object. Hence, as far as the effects of diffraction are negligible, the spatial and angular resolutions are completely decoupled. The required number of correlation events is the same as in chaotic ghost imaging \cite{note_ghost}. We finally remark that the limitations to CPI are ultimately related with fundamental physical constraints (namely, the uncertainty principle) rather than with the geometrical structure of the system.

\begin{figure}
\centering
\includegraphics[width=0.44\textwidth]{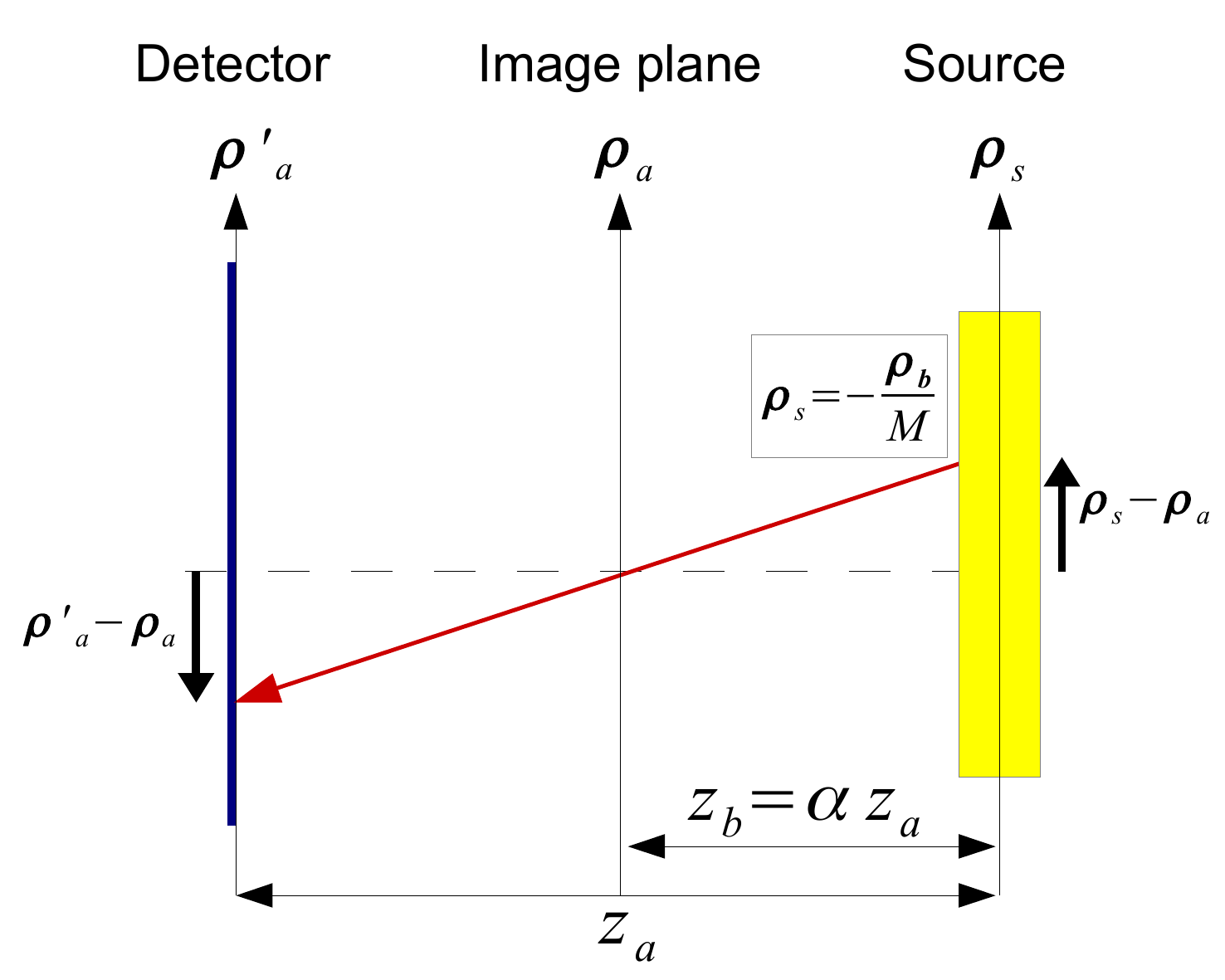}
\caption{Geometrical representation of the scaling property given in Eq.~(\ref{scaling}), in the case $z_a>z_b$.}\label{fig:scaling}
\end{figure}

\begin{figure}[h!]
\centering
\includegraphics[width=0.5\textwidth]{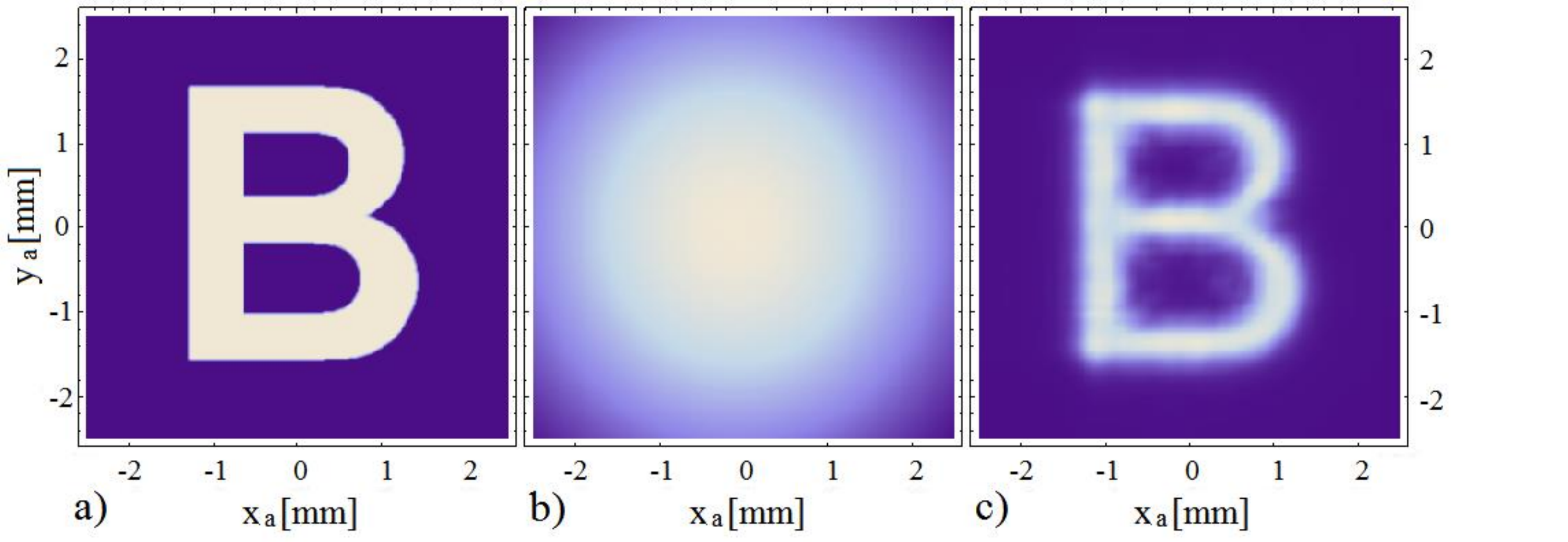}
\caption{Simulations of a CPI system illuminated by a chaotic source with $\lambda=500\,\mathrm{nm}$ and a Gaussian intensity profile of width $D_s \simeq 3\sigma=1.8\,\mathrm{mm}$; the source is magnified by $M=0.8$, the pixel size is $\delta=32\,\mu\mathrm{m}$, the number of pixels for spatial and directional resolutions are $N_x^{(cp)}=N_u^{(cp)}=150$. (a) Focused image in $z_a=z_b=10\,\mathrm{mm}$. (b) Out-of-focus correlation image retrived in $z_a=10\,\mathrm{mm}$, with $z_b=50\,\mathrm{mm}$. (c) Refocused image as given by Eq. (\ref{refocus}).  }\label{fig:refocused}
\end{figure}

In view of exploiting the plenoptic properties of the second order correlation function, let us now consider the more interesting case in which the retrieved image of the scene is out of focus ($z_a\neq z_b$). Inspection of Eq.~(\ref{G12}) indicates that the correlation function $\Gamma_{z_a,z_b}(\bm{\rho}_a,\bm{\rho}_b)$ incorporates a plenoptic information on the lightfield propagating from the source to the object and the sensors, thus enabling the reconstruction of misfocused images. In fact, as schematically shown in Fig.~\ref{fig:scaling}, the propagation direction of the light detected at the point $\bm{\rho'}_a$ of the sensor $\mathrm{D}_a$ can be reconstructed by knowing the source point $\bm{\rho}_s$ from which it is coming. This reconstruction enables re-tracing the light-path from the source to the correct image plane, which is at a distance $z_b=\alpha z_a$ from the source. Imaging the chaotic source on detector $\mathrm{D}_b$ is thus the key feature which enables refocusing in CPI. In fact, from Fig.~\ref{fig:scaling}, one can infer the following scaling property:
\begin{equation}\label{scaling}
\Gamma_{z_a,z_b}\!\left(\frac{z_a}{z_b} \bm{\rho}_a - \frac{\bm{\rho}_b}{M}
\!\left( 1- \frac{z_a}{z_b} \right)\!, \bm{\rho}_b \right) \simeq \Gamma_{z_b,z_b}\!\left(\bm{\rho}_a, \bm{\rho}_b \right),
\end{equation}
which is exact in the geometrical optics limit \cite{note_stat_phase}. Similar to PI \cite{ng}, such a scaling property guarantees the refocusing capability of CPI. Its integral yields the refocused incoherent image:
\begin{equation}\label{refocus}
\Sigma^{\mathrm{ref}}_{z_a,z_b} (\bm{\rho}_a)\! := \!\int \!
d^2\!\bm{\rho}_b \Gamma_{z_a,z_b} \!\left(\frac{z_a}{z_b} \bm{\rho}_a \! -
\frac{\bm{\rho}_b}{M} \!\left( 1- \frac{z_a}{z_b} \right)\!,
\bm{\rho}_b \right). 
\end{equation}  
Figure \ref{fig:refocused} reports the simulation of CPI for: a) a focused image, b) an out of focus image taken at $z_b=5 z_a$, c) the refocused image obtained by applying the scaling rule of Eq.~(\ref{refocus}) to the out-of-focus image. We conclude that the correlation properties of chaotic light \cite{laserphys,ferri} enable achieving a much higher refocusing capability with respect to plenoptic imaging.

The computational steps required for rescaling [Eq.~(\ref{scaling})] and integration [Eq.~(\ref{refocus})] grow linearly in the total number $N_u^2$ of angular pixels. Since the operation must be repeated $N_x^2$ times, the computational steps required for refocusing scale like $(N_u N_x)^2$, as in standard plenoptic imaging. The overall computational time has an additional scaling factor, given by the number of frames which must be averaged to obtain the correlation image.

\textit{Depth of focus.---} It is worth comparing the performance of PI and CPI in terms of the maximum achievable depth of focus (DOF), namely, the maximum distance from the actual detection plane at which perfect refocusing is allowed. For any plenoptic device, we can define $\alpha=S_i/S'_i$ as the ratio between the distance $S_i$ from the focusing element to the image plane, and the distance $S'_i$ between the focusing element (the main lens in standard plenoptic, the source in CPI) and the detector. Perfect refocusing is possible when \cite{ng}:
\begin{equation}\label{ref}
\left| 1-\frac{1}{\alpha} \right| < M \frac{\delta_x}{\delta_u} = \frac{\Delta x}{\Delta u}, 
\end{equation}
where $\Delta x=2\delta_x$ is the minimum distance that can be resolved on the image plane, and $\Delta u=2 \delta_u/M$ is the minimum distance that can be resolved on the focusing element, with $M$ the latter's magnification. Now, in standard PI, the image resolution is given by the width of the {\it macro}pixel $\Delta x^{(\mathrm{p})}= 2 \delta N_u^{(\mathrm{p})}$, while each ({\it micro})pixel $\delta$ coincides with a region on the lens plane of width $\Delta u^{(\mathrm{p})} = 2 D_s/N_u^{(\mathrm{p})}$, hence:
\begin{equation}\label{DOF_p}
\left( \frac{\Delta x}{\Delta u} \right)^{(\mathrm{p})} = \frac{\delta}{D_s} \left( N_u^{(\mathrm{p})} \right)^2.
\end{equation}
In CPI, the relation $\Delta u^{(\mathrm{cp})}=2 D_s/N_u^{(\mathrm{cp})}$ is unchanged, but $\Delta x^{(\mathrm{cp})}=2 \delta$, since pixels of width $\delta$ can be used also to retrieve the image. Hence
\begin{equation}\label{DOF_cp}
\left( \frac{\Delta x}{\Delta u} \right)^{(\mathrm{cp})} =
\frac{\delta}{D_s} N_u^{(\mathrm{cp})}.
\end{equation}
In conclusion, CPI enable to extend perfect refocusing at a much longer distance with respect to PI, since
\begin{equation}\label{refcomp}
\frac{\mathrm{DOF}^{(\mathrm{cp})}}{\mathrm{DOF}^{(\mathrm{p})}} = \frac{N_u^{(\mathrm{cp})}}{( N_u^{(\mathrm{p})} )^2}
\end{equation}
can be easily made larger than one in experiments.

Based on Eq.~(\ref{ref}), in the simulation reported in Fig.\ref{fig:refocused}, the maximum object distance for perfect refocusing is $z_b=5 z_a$. For the given total number of pixels per side $N_{\mathrm{tot}}=300$, a PI system with the same spatial resolution $N_x^{(p)}=150$, would lead to an angular resolution of $N_u^{(p)} = N_{\mathrm{tot}}/N_x^{(p)} = 2$. Based on Eq.~(\ref{refcomp}), the CPI system enables refocusing at a depth of focus that is almost 40 times higher than for the equivalent PI system.

\textit{Conclusions and outlook.---} We have presented an innovative approach to plenoptic imaging which exploits the fundamental correlation properties of chaotic light to decouple spatial and angular resolution of standard plenoptic imaging systems. This has enabled us to perform plenoptic imaging at a significantly higher depth of field with respect to an equivalent standard plenoptic imaging device. As plenoptic imaging is being broadly adopted in diverse fields such as digital photography \cite{website}, microscopy \cite {microscopy2,microscopy4}, 3D imaging, sensing and rendering \cite{3dimaging}, our proposed scheme has direct applications in several biomedical and engineering fields. We have analyzed the CPI setup as an imaging device; potentially, it would also represent an interesting tool to characterize turbulence, thus enabling volumetric imaging within scattering media \cite{waller_turb}. Interestingly, the coherent nature of the correlation plenoptic imaging technique may lead to innovative coherent microscopy modality. The merging of plenoptic imaging and correlation quantum imaging has thus the potential to open a totally new line of research and pave the way towards the promising applications of plenoptic imaging. 

In view of practical applications, it is worth mentioning that the obtained results do not depend on the nature of the object, whether reflective or transmissive. It is also reasonable to expect the CPI procedure to work with any source, of either photons or particles \cite{ghost_muons}, that is characterized by correlation in \textit{both} momentum \textit{and} position \cite{laserphys,gatti2}. In particular, when replacing the chaotic source with an entangled photon source such as Spontaneous Parametric Down-Conversion \cite{klyshko}, the light source can still be imaged on $\mathrm{D}_b$ to obtain the angular information, but a lens is required to achieve ghost imaging of the object \cite{pittman,laserphys}. On the contrary, we do not expect CPI to work with classically correlated beams of the kind employed in Ref.\cite{bennink}, which are only characterized by momentum correlation \cite{laserphys}.

\begin{acknowledgments}
\textit{Acknowledgments.---} The Authors thank O. Vaccarelli for realizing the drawing reported in Fig.\ref{fig:schemes} and for discussions. This work has been supported by the MIUR program P.O.N.~RICERCA E COMPETITIVITA' 2007-2013 - Avviso n.~713/Ric.~del 29/10/2010, Titolo II - ``Sviluppo/Potenziamento di DAT e di LPP'' (project n.~PON02-00576-3333585), the INFN through project QUANTUM, the UMD Tier 1 program and the Ministry of Science of Korea, under the ``ICT Consilience Creative Program'' (IITP-2015-R0346-15-1007).
\end{acknowledgments}


\begin{thebibliography}{10}

\bibitem{adelson}
E.H. Adelson, and J.Y.A. Wang, IEEE Transactions on Pattern
Analysis and Machine Intelligence, \textbf{14}, 99 (1992).

\bibitem{ng}
R. Ng, M. Levoy, M. Br\'edif, G. Duval, M. Horowitz, and P.
Hanrahan, Tech. Rep. CSTR 2005-02, Stanford Computer Science
(2005).

\bibitem{website}
https://www.lytro.com/illum; http://www.raytrix.de/ ; http://www.pelicanimaging.com


\bibitem{3dimaging}
X. Xiao, B. Javidi, M. Martinez-Corral, and A. Stern, Appl. Opt. {\bf 52}, 546 (2013).

\bibitem{waller_turb}
H. Liu, E. Jonas, L. Tian, Z. Jingshan, B. Recht, L. Waller, Opt. Expr. {\bf 23}, 14461 (2015).

\bibitem{muenzel}
S. Muenzel and J. W. Fleischer, Appl. Opt. \textbf{52}, D97 (2013).

\bibitem{levoy}
M. Levoy and P. Hanrahan, \textit{Proceedings of SIGGRAPH '96 (New Orleans, LA, August 4--9, 1996)} pp. 31--42, Computer Graphics Proceedings, Annual Conference Series (ACM SIGGRAPH, New York, 1996).

\bibitem{microscopy1}
M. Levoy, R. Ng, A. Adams, M. Footer, and M. Horowitz, ACM Transactions on Graphics \textbf{25}, 924 (2006). 

\bibitem{microscopy3}
W. Glastre, O. Hugon, O. Jacquin, H. Guillet de Chatellus, and E. Lacot, Opt. Expr. \textbf{21}, 7294 (2013).

\bibitem{microscopy2}
M. Broxton, L. Grosenick, S. Yang, N. Cohen, A. Andalman, K. Deisseroth, and M. Levoy, Opt. Expr. \textbf{21}, 25418 (2013).

\bibitem{microscopy4}
R. Prevedel, Y.-G. Yoon, M. Hoffmann, N. Pak, G. Wetzstein, S. Kato, T Schr\"odel, R. Raskar, M. Zimmer, E. S. Boyden, and A. Vaziri, Nat. Meth. \textbf{11}, 727 (2014).



\bibitem{waller}
L. Waller, G. Situ, and J. W. Fleischer, Nat. Phot. \textbf{6}, 474 (2012).

\bibitem{spatioangular}
T. Georgiev, K. C. Zheng, B. Curless, D. Salesin, S. Nayar, and C. Intwala, \textit{Spatio-Angular Resolution Tradeoff in Integral Photography}, in \textit{Eurographics Symposium on Redering (2006)}, eds. T. Akenine-M\"oller and W. Heidrich (The Eurographics Association, Geneva, 2006).

\bibitem{imageformation}
S. A. Schroff and K. Berkner, Appl. Opt. \textbf{52}, D22 (2013).

\bibitem{superres}
J. P\'erez, E. Magdaleno, F. P\'erez, M. Rodr\'iguez, D. Hern\'andez, and J Corrales, Sensors \textbf{14}, 8669 (2014).



\bibitem{pittman}
T.B. Pittman, Y.H. Shih, D.V. Strekalov, and A.V. Sergienko, Phys. Rev. A {\bf 52}, R3429 (1995); T.B. Pittman, D.V. Strekalov, D.N. Klyshko, et al., Physical
Review A {\bf 53}, 2804 (1996).

\bibitem{gatti2}
A. Gatti, E. Brambilla, M. Bache, and L. A. Lugiato, Phys. Rev. A {\bf 70}, 013802 (2004).

\bibitem{laserphys}
M. D'Angelo, and Y.H. Shih, Laser Phys. Lett. {\bf 2}, 567 (2005).

\bibitem{valencia}
A. Valencia, G. Scarcelli, M. D'Angelo, and Y.H. Shih, Phys. Rev. Lett. {\bf 94}, 063601 (2005).

\bibitem{scarcelliPRL}
G. Scarcelli, V. Berardi, and Y.H. Shih, Phys. Rev. Lett. {\bf 96}, 063602 (2006).

\bibitem{ferri}
F. Ferri, D. Magatti, A. Gatti, M. Bache, E. Brambilla, and L. A. Lugiato, Phys. Rev. Lett. {\bf 94}, 183602 (2005).

\bibitem{mancini}
S. Mancini, V. I. Man'ko, P.Tombesi, Phys. Lett. A {\bf 213}, 1 (1996).

\bibitem{lvovsky}
A. I. Lvovsky and M. G. Raymer, Rev. Mod. Phys. {\bf 81}, 299 (2009).

\bibitem{note_La}
Whenever, for practical purposes, the image needs to be reproduced at a distance from the source different from the object distance $z_b$, a lens must be introduced at a distance $s_i$ from detector $\mathrm{D}_a$ such that $1/(z_a-z_b)+1/s_i=1/f$, where $z_a$ is the distance of the lens from the source and $f$ is the lens focal length \cite{valencia}. This enables displacing the correlation image of the object, obtained at a distance $z_b$ from the source, to any convenient location, without changing the physics and the effectivness of the CPI scheme.

\bibitem{torino}
G. Brida, M.V. Chekhova, G.A. Fornaro, M. Genovese, E. D. Lopaeva, I. Ruo Berchera, Phys. Rev. A \textbf{83}, 063807 (2011).

\bibitem{note_time}
With a sCMOS camera with full well capacity $3\times 10^4$ electrons/s and acquisition time $5\,\mathrm{ms}$ for a $10^6$ pixel image, one needs a photon rate at the camera of $40000\,\mathrm{s}^{-1}$. This can be easily achieved with any pseudo-thermal source made of a laser beam impinging on a moving scattering medium or a spatial light modulator. In the case of low power, a chaotic source such as a LED with wavelength $\lambda=630\,\mathrm{nm}$ and FWHM $20\mathrm{nm}$ needs to be filtered to a FWHM of about $10^{(-10)} nm$ to match the bandwidth of the camera; hence, the required power of the LED is of the order of $1\,\mathrm{mW}$. The constraint on the source bandwidth can be released when high power is available (see Fig.~3 in Ref.~\cite{torino}). Also in this case, even considering collection losses, both pseduo-thermal sources and ultra-high power LEDs available on the market (P = $700\,\mathrm{mW}$) represent realistic sources for implementing CPI. 


\bibitem{scully}
M. O. Scully and M. S. Zubairy, \textit{Quantum Optics} (Cambridge University Press, Cambridge 1997).

\bibitem{goodman}
J. W. Goodman, \textit{Introduction to Fourier Optics} (McGraw-Hill, New York 1996).

\bibitem{note_ghost}
In a realistic situation, such as the one in Figure~\ref{fig:refocused}, with a reasonable source power and camera \cite{note_time}, the total acquisition time is around $25\,\mathrm{s}$ for $5000$ acquisitions for a $10^6$ pixels image, which can go down to about $2\mathrm{s}$ for a $128\times 128$ pixels image. This time can be reduced by implementing compressed sensing techniques (see, e.g., Ref.~\cite{sensing}).

\bibitem{sensing}
Liu Jiying, Zhu Jubo, Lu Chuan, and Huang Shisheng, Optics Letters {\bf 35}, 1206 (2010).

\bibitem{note_stat_phase}
The result of Eq.~(\ref{scaling}) can be formally obtained from Eq.~(\ref{G12}) by considering the stationary-phase approximation of the integrand in the limit $\omega\to\infty$. In fact, in this limit, the most prominent contribution to the integral comes from vanishingly small regions around the points
\begin{eqnarray}
\bm{\rho}_o = \frac{z_a}{z_b} \bm{\rho}_a - \frac{\bm{\rho}_b}{M} 
\!\left( 1- \frac{z_a}{z_b} \right), \qquad \bm{\rho}_s = - \frac{\bm{\rho}_b}{M}, \nonumber
\end{eqnarray}  
thus justifying the intuitive result of Eq.~(\ref{scaling}).  

\bibitem{ghost_muons}
M. D'Angelo, A. Garuccio, F. Romano, F. Di Lena, M. D'Incecco, R. Moro, A. Regano, G. Scarcelli, Springer Proceedings in Physics \textbf{145}, 237 (2014)

\bibitem{bennink}
R.S. Bennink, S.J. Bentley, and R.W. Boyd, Phys. Rev. Lett. {\bf 89}, 113601 (2002).



\bibitem{klyshko}
D. N. Klyshko, \textit{Photons and Nonlinear Optics} (Gordon and Breach, New York 1988).

\end{thebibliography}
\end{document}